\documentclass[letterpaper]{article}
\usepackage{graphicx}
\usepackage{amssymb}
\usepackage{amsmath}
\usepackage{amsfonts,enumerate}
\begin{document}

\title{Privacy Preserving and Ownership Authentication in Ubiquitous Computing Devices using Secure Three Way Authentication}
\author{Pradeep B.H and Sanjay Singh\thanks { Sanjay Singh is with the Department of Information and Communication Technology, Manipal Institute of Technology, Manipal University, Manipal-576104, India, E-mail: sanjay.singh@manipal.edu}}

\maketitle
\begin{abstract}
In todays world of technology and gadgets almost every person is having a portable device, be it a laptop or the smart phones. The user would like to have all the services at his fingertips and access them through the portable device he owns. Maybe he wants some data from the fellow user or from the service provider or maybe he wants to control his smart devices at home from wherever he is. In the present era of mobile environments, interactions between the user device and the service provider must be secure enough regardless of the type of device used to access or utilize the services. In this paper we propose a "Secure Three Way Authentication (STWA)" technique intended to preserve the user privacy and to accomplish ownership authentication in order to securely deliver the services to the user devices. This technique will also help the users or the service providers to check if the device is compromised or not with the help of the encrypted pass-phrases that are being exchanged.
\end{abstract}

\section{Introduction}
A ubiquitous computing (Ubicomp) or pervasive computing environment is imagined as a system with large number of invisible, collaborating computers, sensors and actuators interacting with the user devices. Data about individuals who are in the environment is constantly being generated, transmitted, modified and stored. The user data present in the environment will be very sensitive. Protecting private data of every user is a major concern. Also in the this era of the mobile environment the user owns a number of portable devices like the PDAs, Laptops, Mobile Phones etc. with varying computing capabilities in order to access the variety of services that are being provided by the service providers. It is very much important to secure the service interactions between the user and the service providers. If the interactions or the transactions are not secure then the user will be hesitant to avail the services by providing the most sensitive data hence revenue loss for the service providers. For example a user who wants to have a secure bank transaction will not go in for accessing his/her account by providing the username, password and also his/her account details if he/she is not sure if the connection is secure. Hence it is important that the user's details are hidden from the third party and provide security. Previously many approaches have been proposed in this regard. 
\par
Jalal Al-Muhtadi et al \cite{jard} suggested different wearable and embedded devices such as smart jeweleries, active badges and smart watches etc which contain an ID for authentication, but the user should carry it wherever he goes. Also there are chances of the device being lost and fall into wrong hands. U.P.Kulkarni et al \cite{kvjcy} and C. Lesniewski et al \cite{cbjrf} has used the concept of  Certifying Authority for authentication, which requires the user to register his/her devices and also requires the user to maintain his/her certificate timely. A non technical user would find it difficult to manage the certificates and it would be a burden. If a person is having more than one device, then he/she needs to have independent certificates for all the devices he/she possesses and should be managing his/her certificate devices time to time and should be spending more time on this rather than actually doing his/her work. 
\par
Wenjuan Liu et al \cite{wxsjs}, has used the concept of information hiding used in TCP/IP packets. However this approach might not be useful at all time. It can mostly be used as the trustworthy authentication of security devices such as fire walls. Every time the information or the request is being sent, it will be encapsulated. This may lead to the encapsulation of non-sensitive information. The encapsulation will not be able to differentiate between the sensitive and non-sensitive information. This limitation will lead to the high computational and transmission overhead. Any loss in the data during transmission will lead to the inconsistency in the request or the information sent.
\par
 Adrian Leung and Chris J. Mitchell \cite{ac} in their work have proposed manual authentication protocol to authenticate the user and his/her device. This protocol authenticates the two devices using a combination of an insecure wireless channel and manual data transfer. As it uses the insecure channel, the system may be susceptible to attacks of any kind for the wireless network \cite{lis}. Whatever is the information, be it sensitive information or non-sensitive can be tracked and attacked. Moreover the user has to transfer the data manually in order to be authenticated. In the present era people tend to go towards automation of the systems. Therefore it will not be good enough to go with the manual system for authentication. Also the author mentions that the user, user device and the device management entity needs to be in close physical proximity with each other during initialization phase, which will actually wipe out the concept of mobility and ubiquitous computing.
\par
In this paper we have proposed an idea of Secure Three way Authentication technique which provides a better security during the service interactions between the users and the service providers. It also addresses most of the limitations mentioned above.
\par
Rest of the paper is organized as follows. Section II explains the operation of the proposed three way authentication technique followed by discussion in section III. Finally section IV concludes the paper.

\section{Secure Three Way Authentication (STWA)}
\label{a}
In this paper we propose a simple and more user friendly approach to authenticate the user and his/her device in the ubiquitous environment. 
\begin{table}[bpht!]
\centering
	\caption{Notations Used} 
	\label{tab}
	\begin{tabular}{|c|p{2in}|c|p{2in}|}
	\hline 
	\textbf{Symbol}&\textbf{Meaning}&\textbf{Symbol}&\textbf{Meaning} \\ \hline 
		$T_A$&Token of the corresponding user A signed by the KDC&$CKS$&Central Key Server\\ \hline 
			$ID_{machine}$& Machine ID or the Device Serial Number&$KDC$&Manufacturer's Key Distribution Center \\ \hline 
			$U_A$&User A&$U_B$&User B \\ \hline
			$U_M$&Device Manufacturer& $DMN$&Device Model Number\\ \hline
			$E_{P_{CKS}}$&Encryption Using Public key of CKS&$E_{P_{KDC}}$&Encryption Using Public key of KDC\\ \hline
			$N_A$&nonce generated by A& $N_B$&Nonce generated by B \\ \hline
			$N_U$&Nonce generated by user U& $N_{CKS}$&Nonce generated by CKS \\ \hline
			$ID_B$&User ID or User Name of the user B& $ID_A$&User ID or User Name of the user A\\ \hline 
			$ID_M$&Manufacturer's ID&$T_S$&Time Stamp\\ \hline
			$P_{CKS}$&Public key of the CKS& $PW_U$&Password of the User U\\ \hline
			$P1$&Pass-phrase which is known to central key server alone&$P2$&Pass-phrase which is known to user and the CKS \\ \hline
			$K$&One Time Session Key&$SR$& Service Request\\ \hline
			$SP$& Service Provider&$Ack_D$& Acknowledgment for device authentication at the time of registration\\ \hline
		  $Ack_U$& Acknowledgment for user registration&$H$& Hash of the message using MD5 or SHA1 etc.\\ \hline
		  $OTP$&One Time Password&$TempID$&Temporary Identity or pseudo Identity or the respective user \\ \hline
		  $N_M$&Nonce of manufacturer&$\|$&Concatenation Symbol\\ \hline
		  
	\end{tabular}
\end{table}	 
The proposed solution has the following phases:
\begin{itemize}
\item Initialization
\item	User Registration
\item	Connection between users 
\item	High Level Transaction
\end{itemize}

\subsection{Initialization}
This phase is carried out at the site of the manufacturer and the manufacturer is treated as the user here. The Assumption made is that the manufacturer is already an authenticated person. The manufacturer will be registering the device with Key Distribution Center (KDC) meant only for the manufacturers by sending the machine ID and the device model number (DMN). The machine ID and the DMN is encrypted by public key of the KDC. The KDC will then generate a token T which is the hash of manufacturers ID and a time stamp. This token T is encrypted using the nonce of the manufacturer and sent to the manufacturer through a secure channel. The token will also be stored in the central key server (CKS) for future use. This is performed before the device is sold. The token obtained by the manufacturer will be encrypted using RSA \cite{huth} and then hard coded into the trusted portability module \cite{tcg}\cite{tpm} which is embedded in the portable device. The message exchange that takes place between the KDC and the manufacturer during the process of initialization is as follows:
			
			\begin{enumerate}
				\item $U_M\rightarrow KDC\ :\ E_{P_{KDC}} (ID_{machine} \| DMN\| N_M\| ID_M)$\\ 
				The manufacturer will send the unique machine ID along with the device model number and nonce ($N_M$) and the ID of the manufacturer to the KDC to register the device being manufactured. This message is encrypted using the public key of the KDC.  
				\item $KDC \rightarrow U_M \ :\ E_{N_M}(T) $, where $T = H(ID_M \| T_S)$ \\ 
				The KDC decrypts the message received using its private key and generates a token T which is a hash of manufacturer's ID and a time stamp which denotes the time when the device was registered. In future the token will be helpful to recognize to which manufacture the device belongs to. Then the KDC sends the token T through secure channel by encrypting it using the nonce of the manufacturer.  
				\item $KDC \rightarrow CKS \ :\ E_{P_{CKS}} (ID_{machine} \| DMN \| T)$\\ 
				Once the token has been generated and sent to the manufacturer, the work of the KDC is done. Now KDC will send a encrypted message to CKS using the server's public key. The message consists of the machine ID, device model number and the token T. This is done because the CKS needs this information to carry out the device authentication in future transactions.   
			\end{enumerate}
			
\begin{figure}[bpht!]
\centering
\includegraphics[width=3in,height=2in]{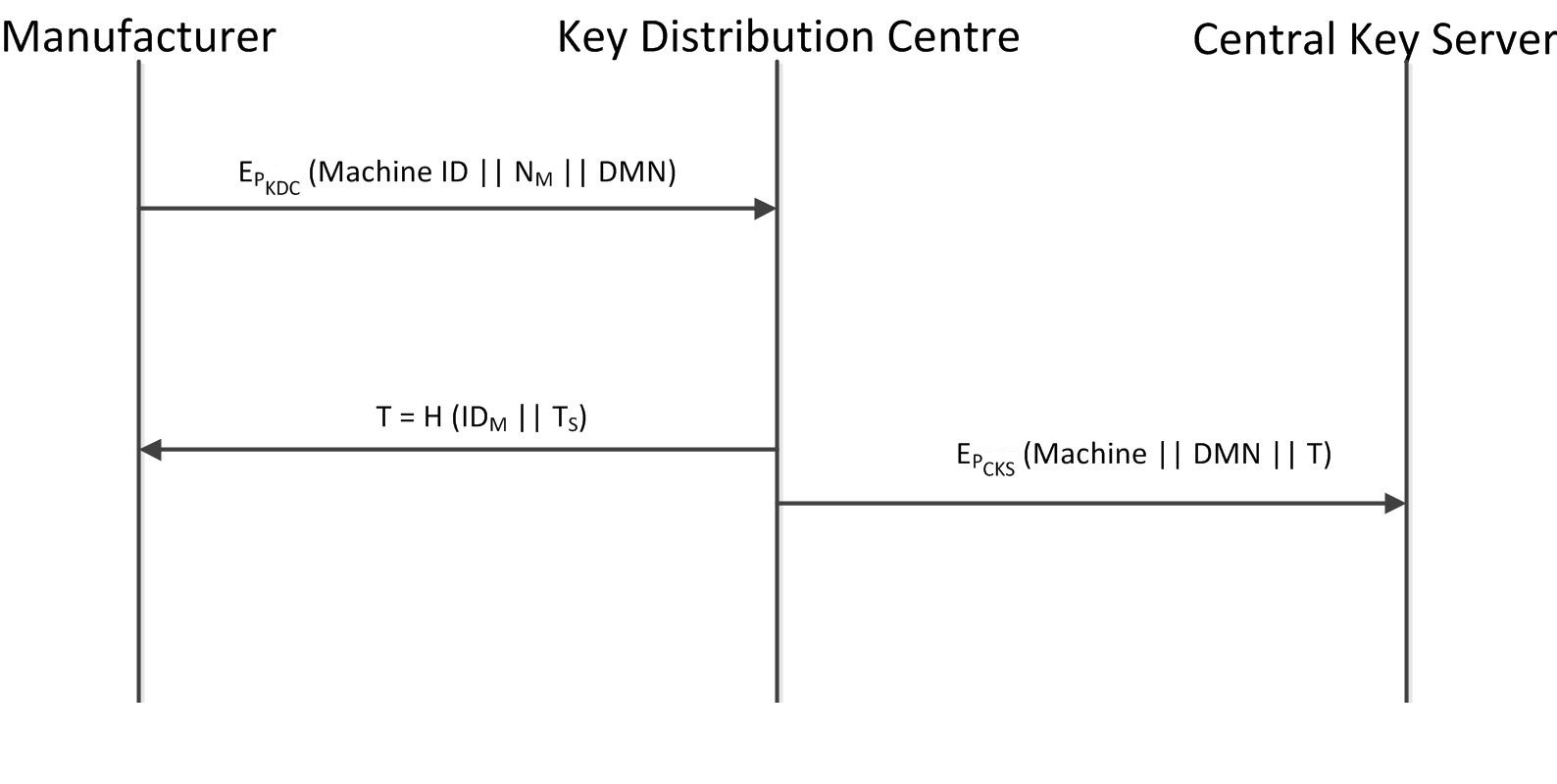}
\caption{Diagram Showing Device Initialization Process.}
\label{fig:1}
\end{figure}

The above explained process of device initialization is summarized in Fig. \ref{fig:1}.

\subsection{User Registration}
This phase is carried out only after the device is sold. The user who is the owner of the device needs to register to the CKS. He/she sends the message containing the machine ID and the token encrypted using the public key of the CKS; the CKS will decrypt the message received from the user by its private key. The token and the machine ID sent by the user is compared with the machine ID and the token which has been stored in the server's database during the initialization phase. If both are same then the server sends the acknowledgment saying the authentication of the device was successful and also an one time password (OTP). Now the user needs to register himself/herself by providing the user ID and the OTP to CKS. Once the registration is done, the server sends an acknowledgment to the user along with the Temp ID. Now the user is ready to avail any service through his/her device. The messages that are exchanged during this phase is given below.
				\begin{enumerate}
					\item $U \rightarrow CKS \ :\ E_{P_{CKS}}(ID_{machine} \| T \| N_U)$\\ 
					Once the device is bought, the use needs to register himself/herself to the CKS. To do so the device has to be authenticated. So the user sends a message containing the machine ID, $N_U$ and the token T, encrypted using the public key of the CKS. 
					\item CKS decrypts the contents of the message using its private key and compares them with the contents in its database. Only when the contents match, the CKS will send the success message. 
					\item $CKS \rightarrow U\ : \ E_{N_U}(Ack_D \| OTP)$\\ 
					CKS will send the device acknowledgment $(Ack_D)$ saying the device has been successfully authenticated, which indicates the user that he needs to register him/her to the CKS using the one time password (OTP) received by him/her where $Ack_{D} = H( T \| DMN )$. This whole message is encrypted by using the nonce of the user.
					\item $U\rightarrow CKS\ :\ E_{P_{CKS}} (ID_U \| OTP\|N_U)$\\  
					Now the user sends a message containing his credentials like User ID, OTP and a nonce. The message is encrypted using the public key of the server. A nonce is like one time password which is generated by the device itself for every new transaction. When the nonce is generated, a corresponding pair which acts as a private key also has to be generated for decrypting the encrypted message.
					\item $CKS \rightarrow U \ :\ E_{N_{U}}(Ack_U\| Temp ID)$ \\ 
					Once the message from the user is received by the CKS, it will decrypt it using its private key and register the user. At the end of this phase, the server sends an user registration acknowledgment $(Ack_U)$ and a TempID which is encrypted using the nonce sent by the user. The user after receiving the message will decrypt it and retrieve his/her TempID.
				\end{enumerate}
				
\begin{figure}[bpht!]
\centering
\includegraphics[width=3.25in,height=3in]{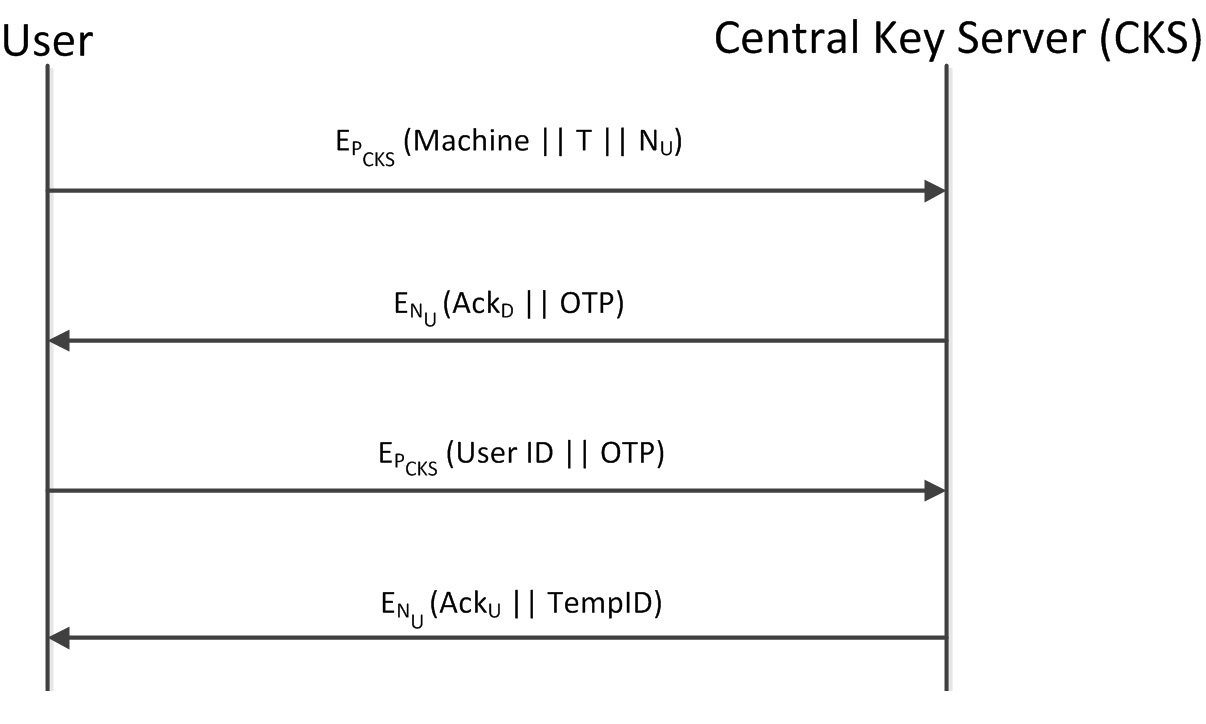}
\caption{Diagram Showing the User Registration Process.}
\label{fig:2}
\end{figure}

The process of user registration is shown in Fig. \ref{fig:2}.

\subsection{Connection between the Users}
\begin{figure}[bpht!]
\centering
\includegraphics[height=10cm, width=12cm]{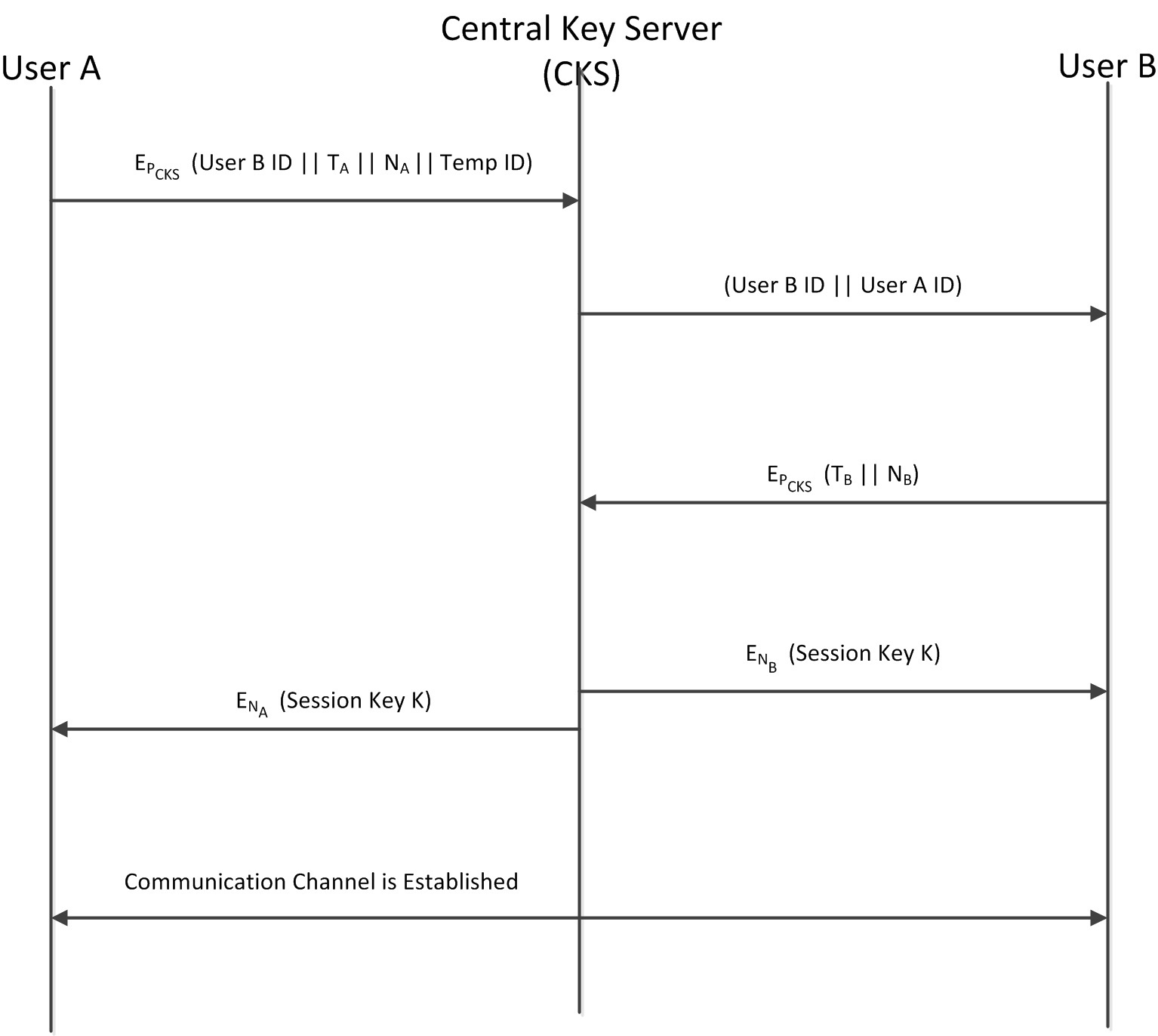}
\caption{Diagram Showing Connection Between the Users.}
\label{fig:3}
\end{figure}

Every user will be given a TempID from the CKS during the user registration phase. In this phase and in the high level transaction phase, user needs to send his/her TempID along with the nonce that is generated in the device. To connect the device to the fellow user's device, the user needs to connect through the Internet. The user, who wants to connect to a particular user, sends the token and his/her Temp ID along with the details of the device he wants to connect, to the CKS in the network. The key server compares the token and his/her ID with that present in its database. If they are same, then it requests the called user to send his/her details and compares it. If the called user credentials are verified, then the CKS sends a one time session key to both the devices. If any of the user's details is not matching with his/her details in its database then the server sends the message to the other user that the device is not an authorized one and he/she is trying to connect to the malicious entity. 
The scenario of the connection between the users described is as follows.
\begin{enumerate}
			\item $U_A \rightarrow CKS\ :\ E_{P_{CKS}}(ID_{U_B} \| T_A\| N_A\|TempID)$\\
			 The user who is interested in establishing a connection with the other user of his interest sends the target user ID along with his token, a nonce and his/her TempID. This message is encrypted using the public key of the CKS.
			\item $CKS \rightarrow U_B\ : (ID_B \| ID_A)$\\ CKS decrypts the message received and compares it with the contents in its database. It check for the users details in its database. If found, the user is authenticated. It also checks for the device details and if they match with the details in its database, then the CKS sends the message to the user saying that the user A is trying to connect to user B.
			\item $U_B \rightarrow CKS \ :\ E_{P_{CKS}}( T_B\| N_B)$\\ Once the message from the CKS is received by the user B, he/she will send the details like token and a nonce by encrypting them using the CKS's public key.
			\item  CKS decrypts the message sent by the User B. If authenticated, both the users will be given a one time session key ($K$) to encrypt and decrypt the messages that are being exchanged between them. There should be some time bound for the session keys, after which it must get expired by itself. It will avoid the replay attack. The session key sent to both the users will be encrypted using their respective nonce. \\ 
			$CKS\rightarrow U_A\ : \ E_{N_A}(K)$\\
			$CKS \rightarrow U_B\ : \ E_{N_B}(K)$\\
		\end{enumerate}
The process of connection between the users is summarized in Fig. \ref{fig:3}.

\subsection{High Level Transactions}
The user does not just try to connect his/her devices to the other users devices, but goes for the transactions wherein the user needs to submit his most important and sensitive details to the service providers. For example he/she may try to access his bank accounts from his/her device where he/she needs to submit his/her account details, this kind of transactions are generalized as high level transactions. The user who wants to have a high level transaction will send the request to the central key server to grant access to a particular service provider. This technique is analogous to Kerberos, where the central key server (CKS) acts as both the authentication server and the ticket granting server. The service provider is the server to which the user wants to gain access. In case of the bank transaction there is no need to search for the server as a bank has a dedicated server. In case of other transactions like online shopping the service providers are more than one and the best service provider has to be chosen. In this technique two different pass-phrases (P1 \& P2) are used. P1 is what the central key server knows and P2 is what the central key server and the user knows. The pass- phrase P2 is entered manually by the user and pass-phrase P1 is generated by the central key server. The pass-phrases act as a secret key or word adding up an extra layer of security in ubiquitous computing. The proposed technique can be explained in different steps as follows: 

\begin{enumerate}
		\item $U\rightarrow CKS\ :\ E_{P_{CKS}}(SR\| P2 \| T\| N_U\|TempID)$\\ The user sends a service request, pass-phrase P2, token, TempID and nonce encrypting it using the public key of the CKS. 
		
		\item CKS decrypts the message and compares the contents with that in the database and saves P2. If the user is authenticated, the CKS will check the service request from the user and sends the ID of the service provider who is best suited to provide the service that has been requested by the user.
		\item $CKS \rightarrow U \ : \ E_{N_U}(K\|E_{N_{CKS}}(P1)\|ID_{SP}\| P2)$\\
		 CKS will send a session key, a pass-phrase P1 encrypted using the nonce of the CKS and ID of the service provider. This message is encrypted using the nonce of the user. The user cross checks the pass-phrase P2 and if it matches with what it has send in step 1, then it stores the session key. 
		\item $U\rightarrow SP\ : \ ID_{SP}\| E_{N_{CKS}}(P1)$\\ Once the user receives the message, he/she decrypts it. The user sends the message containing ID of the service provider and the encrypted pass-phrase P1 to the service provider. 
		\item $SP\rightarrow CKS\ : \ E_{N_{CKS}}(P1)\|E_{P_{CKS}}(N_{SP}\|ID_U)$\\ The service provider forwards the encrypted pass-phrase P1 along with its nonce and the ID of the user from whom the message was sent by encrypting it using the public key of the CKS to CKS.
		\item CKS checks for the contents of the message with that in its database, only if they match the service provider is authenticated.
		\item $CKS\rightarrow SP\ : \ E_{N_U}(P2\|K\|ID_{SP})\|E_{N_{SP}}(K\|ID_U)$\\ CKS nows sends the pass-phrase P2, session key, and the ID of the service provider encrypted using the nonce of the user along with the other message containing the session key and the user ID encrypted using the nonce of the service provider. 
		\item $SP \rightarrow U \ : \ E_{N_U}(P2\|K\|ID_{SP})$\\ The service provider forwards the pass-phrase P2, session key, and the ID of the service provider encrypted using the nonce of the user, to the user.  
		\item User decrypts the message from the SP, and checks P2, if it is the same what he/she had sent to the CKS and also checks whether the session key is the same as that received earlier. If both P2 and K match with their respective values, the user is now confident that the service provider is not a malicious entity.
		\item Once the User and the service provider get the session key, they will exchange the messages which are encrypted using the session key.\\
		$U\rightarrow SP\ :\ E_K(Message)$\\
		$SP\rightarrow U\ :\ E_K(Message)$\\ 
	\end{enumerate}

\begin{figure}[bpht!]
\centering
\includegraphics[width=3.5in,height=3in]{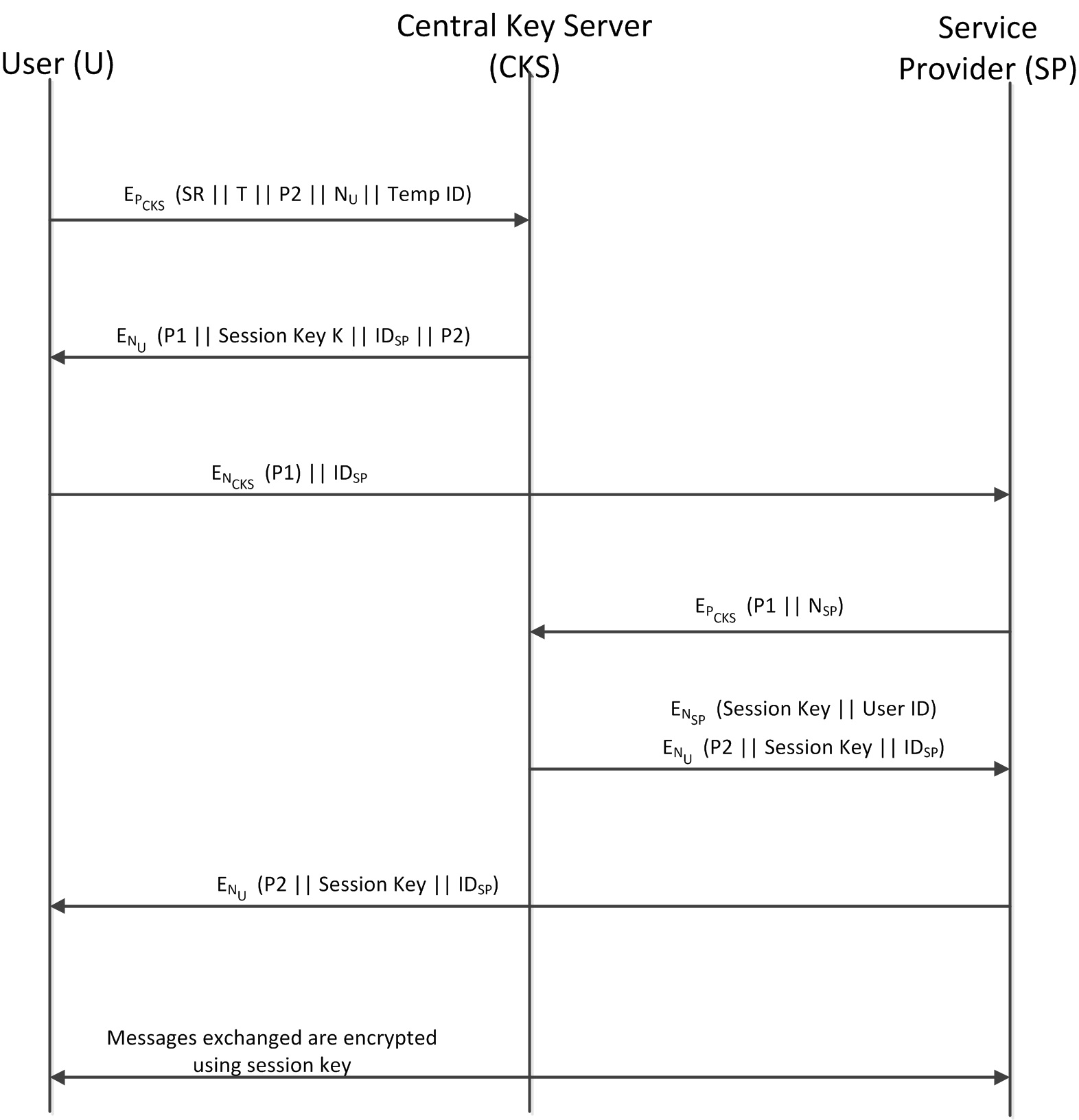}
\caption{Diagram Showing the Process of High Level Transactions.}
\label{fig:4}
\end{figure}

The process of high level transaction is summarized in Fig. \ref{fig:4}.

\section{Discussion}
The Ubicomp devices can establish different types of connections as mentioned in the earlier part of the paper. The device goes through the phase of initialization where the machine ID is registered with the KDC to make sure that the malicious device does not try to access the services or the information from the other users. In the user registration phase the device is first authenticated and then the user is prompted for registering himself/herself to the central key server. This phase adds up to the security aspect as the user with a non-authenticated device will not be allowed to register unless his/her device is authenticated. During phase of connection between the users, the devices are authenticated by their respective tokens. The devices read the exchanged tokens which has the signature of the KDC which helps them in mutual authentication. More over the users submit their TempID's and other credentials which will help the CKS to authenticate the users. The concept of nonce and session key is used which adds up the security. By making use of the nonce the burden on the server to maintain the public keys of all the users and the service providers associated with it is avoided. 
\par
In the high level transaction phase, the two pass-phrases used are known only to the user and the central key server. However the crucial information which the user shares with the central key server is not available to the service provider or to any third party. From this technique the user will be assured that the service provider is also a trusted entity and so the service provider will be having the confirmation that the user who is requesting for its service is not the malicious entity. Since the central key server is the entity in between the user and the service provider, it will also be sure that the true user will be establishing the connection with the trusted service provider. In this technique the user will be accessing the service with the help of his/her Temp ID and his/her actual details will not be available to the service provider. So this technique will protect the privacy of the user and will not allow illegal sharing of user's sensitive data. The technique will also make sure that the compromised device does not get into contact with the service provider. This is done with the help of the pass-phrases. So the Secure Three Way Authentication technique will be a handy technique in the ubiquitous environment.
\par 
A malicious entity between the two users may tap their messages and send the same messages to the server, impersonating to be one among the two users. The malicious entity sends a number of SYN messages to the server. After receiving the SYN-ACK message from the server, the malicious entity will not send the expected ACK message to the server. The malicious entity sends a number of connection requests to the server and tries to create a bottle neck in the network. This is called the SYN flood attack \cite{rfcsyn}. SYN flood attack is a type of denial of service attack where the TCP Three way handshake is not completed as the client will not send the expected ACK messages. The technique presented in this paper avoids the SYN flood attack to a certain extent. As mentioned above, the messages exchanged between the users are encrypted using a one time session key. Since the messages exchanged between the users are in encrypted form, the malicious entity will not be able to know the contents of the messages even though he/she has the access to the encrypted messages. Moreover the messages are exchanged through the server, only the server will know to whom the particular message has to be forwarded and from which user the message has been transmitted. The messages are of no use to the malicious entity as he is not aware of the content. Even though the malicious entity attacks the network by SYN flood attack, it can be avoided by using SYN cookies which eliminate the resources allocated on the target host. As this paper deals only with the user privacy and authentication, more detailed study and analysis with respect to the attack on the network will be carried out as part of our future work.
\vspace{0.1in}
\section{Conclusion}
The security in ubiquitous computing along with authentication and preserving the user privacy is more important in the present world wherein people often get their work done at anytime and from anywhere through their portable devices. 
\par
This paper proposes a new technique of authentication which ensures the privacy of the user and makes sure that the user is convinced about the security he/she is looking for in order to submit the most sensitive data with the intention to avail the services from the service providers. This technique authenticates the user and the service provider to the central key server and also the user and the service provider are mutually authenticated with the help of the central key server. There is no burden on the non technical users to maintain his/her certificates or need for the user to carry authentication devices with them. The future work of the proposed solution would be the inspection of device connections like a social network, which brings to mind the trust management, where the trust quotient of the device should be calculated regularly and quickly. Since this paper concemtrates only on the user privacy and authentication, the attacks on the network, for example Syn flood attack will be dealt as future part of our work.  

\bibliographystyle{unsrt}
\bibliography{ref}  
\end{document}